\documentstyle[psfig]{article}
\begin{document}
\begin{center}
{\large
{\bf Dipole Interactions and Electrical Polarity in Nanosystems --
the Clausius-Mossotti and Related Models}}\\
\vskip 5mm
Philip B. Allen\\
\vskip 3mm
Department of Applied Physics and Applied Mathematics\\
Columbia University, New York, NY 10032\\
and\\
Department of Physics and Astronomy\\
State University of New York, Stony Brook, NY 11794-3800\\
(permanent address)\\
\vskip 5mm
{\bf Abstract}\\
\end{center}
\vskip 5mm \noindent Point polarizable molecules at fixed spatial
positions have solvable electrostatic properties in classical
approximation, the most familiar being the Clausius-Mossotti (CM)
formula.  This paper generalizes the model and imagines various
applications to nanosystems.  The behavior is worked out for a
sequence of octahedral fragments of simple cubic crystals, and the
crossover to the bulk CM law is found.  Some relations to fixed
moment systems are discussed and exploited.
The one-dimensional dipole stack is
introduced as an important model system.  The energy of
interaction of parallel stacks is worked out, and clarifies the
diverse behavior found in different crystal structures. It also
suggests patterns of self-organization which polar molecules in
solution might adopt.  A sum rule on the stack interaction is
found and tested.  Stability
of polarized states under thermal fluctuations is discussed, using
the one-dimensional domain wall as an example.   
Possible structures for polar hard ellipsoids are considered.
An idea is formulated for enhancing polarity of nanosystems
by intentionally adding metallic coatings.

%\pagebreak

\vskip 0.7cm

\section{\label{sec:int}Introduction}

Electrical polarity (neutral charge distributions with
positive and negative charge centers separated in space),
and in particular dipole moments, are
ubiquitous in asymmetric molecules, and therefore no doubt
very common in nanocrystals \cite{Guyot}.  Systematic studies of polarity of
nanocrystals are rare.  Two primitive questions arise.  (1) When
can a nanosystem develop a spontaneous polarity, analogous to a
bulk ferroelectric, i.e. reversible in a strong applied field, and
how can this polarity be controlled \cite{Landauer}?  
A particularly fascinating
example is the recent discovery \cite{Moro} of spontaneous
polarization in small clusters of Nb atoms at low temperature $T$.
(2) What ``permanent'' moments occur, how stable are they in time,
and how much fluctuation occurs within an ensemble of similarly
prepared nanocrystals?

The venerable ``Clausius-Mossotti model'' (CM model, or CMM) 
\cite{Clausius,Mossotti,Aspnes}, provides a simple
solvable picture of spontaneous polarity.  Most of this paper
consists of imagining possible applications of the CMM and working
out some consequences in nanosystems.  I define the CMM to mean a
system of fixed points $\vec{r}_i$, each of which has a
polarizability $\alpha_i$.  When an electric field $\vec{F}_{i,{\rm tot}}$
appears at point $i$, it induces an electric dipole moment
$\vec{\mu}_i = \alpha_i\vec{F}_{i,{\rm tot}}$.  Each such moment creates a
dipole field. The sum of all such dipole fields is the induced field
\begin{equation}
\vec{F}_{i,{\rm ind}}=\sum_j^{\neq i} 
\frac{3\vec{R}_{ij}(\vec{R}_{ij}\cdot\vec{\mu}_j) 
- R_{ij}^2 \vec{\mu}_j}{R_{ij}^5}
\label{eq:Find}
\end{equation}
and the total field is the sum of these plus the
external field $\vec{F}_{i,{\rm tot}}=\vec{F}_{i,{\rm ext}}+\vec{F}_{i,{\rm ind}}$.
This problem requires self-consistency to find the answer for $\vec{\mu}_i$.

The limitation to induced moments and
dipole-dipole interactions inhibits application of
the CMM to real systems.  However, by focusing 
on the dipole-dipole interaction, this model 
clarifies a difficult aspect of real polar systems,
including systems with fixed moments.

There is a large chemical literature developing and
using model interactions between molecules,
in which some or all of the quantum
electron behavior can be represented approximately as classical
charges and dipoles which interact electrostatically
\cite{Dugourd,Rubio,Goddard,Friesner,Madden}
Ideas of this kind are particularly appropriate
for interacting organic molecules ({\it e.g.}
molecular crystals \cite{Silinsh,Rohleder,Tsiper}).
Interactions alter their electronic properties 
from values found in isolated molecules; the effect is
large, but not so large as to destroy the identity of the
molecule as happens in covalent, ionic, or metallic interactions.
The alterations arise from the polarization of surrounding
molecules and can be modelled by classical electrostatics.
The present paper is motivated by more primitive 
questions; the methods developed here can be found
in related but more sophisticated contexts elsewhere.

\section{\label{sec:cml}The ``Generalized Clausius-Mossotti Law'' (CML)}

The textbook version of the CML considers an infinite periodic
medium with polarizable molecules sitting on equivalent sites of
cubic symmetry.  The result is that the total field $\vec{F}$ at
any molecular site is enhanced from the value $\vec{F}_{\rm ext}$
of the homogeneous applied field by an enhancement factor
$f=(1-4\pi n \alpha/3)^{-1}$, where $n$ is the number density of
the molecules.  Equivalently, $\alpha$ is enhanced to the value
$f\alpha$ by the fields of the other dipoles.
A general derivation from microscopic response theory was given
by Maksimov and Mazin \cite{Maksimov}.  

A good example is crystalline C$_{60}$ which
has weakly-coupled molecules of known \cite{Ballard,Pederson} polarizability
$\alpha$=85$\AA^3$ sitting on a face-centered cubic (fcc) lattice of
density $n$=0.001406 molecules/$\AA^3$.  The enhancement factor is
$f=2.00$ and the dielectric constant \cite{Eklund} is 
$\epsilon_{\infty}=1+4\pi n f \alpha$ = 4.0.  
These independently measured values of the
the vapor-phase property \cite{Ballard} $\alpha$, and
bulk-solid property \cite{Eklund} $\epsilon_{\infty}$, 
provide an excellent confirmation of the CML.
Metallic screening by intermolecular charge
transfer, while needed in a complete theory, is small enough
in C$_{60}$ that the Clausius-Mossotti approach works well.

The microscopic calculation of Pederson and Quong
\cite{Pederson} shows that, for C$_{60}$, the theoretical 
free-molecule polarizability ($83.5 \AA^3$) is strongly
affected by intramolecular screening.  Even within
a single molecule the local field is reduced by a factor of
three by intramolecular charge rearrangement.  But when
assembled into the fcc solid, the details of local charge are
successfully ignored by the CMM which places all the
polarity at a single point of cubic symmetry.  There is
no guarantee that this would be equally successful for
less symmetric molecules.  Schemes for representing
the quantum electrostatics of molecules by interacting
``submolecules'' have been developed \cite{Silinsh,Rohleder,Tsiper}.

Note that the CM enhancement factor $f$ diverges to infinity when the
polarizability increases to $\alpha_{\rm max} =3/4\pi n$, or
alternately, the density increases to $n_{\rm max}=3/ 4\pi\alpha$.
There is no known molecular crystal where this limit is reached,
but C$_{60}$ comes within a factor of 2.  If the limit were reached,
the material would enter a phase of spontaneous polarization.
Actual solids which do polarize spontaneously \cite{Jona,Lines} are more
complicated, but the CMM provides a useful simple system with
the same property.

To derive a more general result, note that the dipole moments
$\{\vec{\mu}_i\}$ or their Cartesian components
$\{\mu_{i\alpha}\}$, obey a minimum principle.  That is, they
minimize the energy expression
\begin{equation}
{\cal E}_{\rm dip}\left(\{\vec{\mu}_i\}\right)
=\sum_i \left(\frac{\mu_i^2}{2\alpha} -\vec{\mu}_i \cdot 
(\frac{1}{2}\vec{F}_{i,{\rm ind}}+\vec{F}_{i,{\rm ext}}) \right).
\label{eq:Etot} 
\end{equation}
The first term expresses the fact that it costs less to polarize a
molecule when its polarizability is large.  The factor of $1/2$ in
the second term avoids double counting the dipole-dipole interaction.
The energy formula is a quadratic form, amenable to methods of
linear algebra.  It is helpful to use a vector space notation,
\begin{equation}
{\cal E}_{\rm dip}\left(\{\vec{\mu}_i\}\right)
=\frac{1}{2}<\mu|\frac{1}{\alpha} \hat{\bf 1} - 
{\bf{\Gamma}}|\mu> -<\mu|F_{\rm ext}>,
\label{eq:Emat} 
\end{equation}
where the $3N$-dimensional column vectors have been introduced:
\begin{equation}
|\mu>=\left(
\begin{array}{c} \mu_{1x} \\ \mu_{1y} \\ \vdots \\ \mu_{Nz} 
\end{array}\right) \ \ \ \
|F>=\left(
\begin{array}{c} F_{1x} \\ F_{1y} \\ \vdots \\ F_{Nz} 
\end{array}\right).
\nonumber
\label{eq:cvecs}
\end{equation}
The $3N \times 3N$-dimensional matrix $\Gamma$ is the 
dipole-dipole interaction, defined by Eq.(\ref{eq:Find}), {\it i.e.}
$|F_{\rm ind}>=\Gamma|\mu>$, and has elements
\begin{equation}
\Gamma_{i\alpha,j\beta}=\frac{3R_{ij\alpha}R_{ij\beta}-
\delta_{\alpha\beta}R_{ij}^2 } {R_{ij}^5},
\label{eq:gamma}
\end{equation}
where $R_{ij\alpha}$ is the $\alpha$ Cartesian component of the
vector $\vec{R}_{ij}=\vec{R}_i-\vec{R}_j$.
The solution for the dipoles $\vec{\mu}_i$ which minimize the energy is
\begin{equation}
\vec{\mu}_i=\sum_j \left( \frac{1}{\alpha} \hat{\bf 1} - {\bf{\Gamma}} 
\right)^{-1}_{ij} \vec{F}_{{\rm ext},j} 
\label{eq:cml}
\end{equation}
This is the generalized Clausius-Mossotti law (CML), valid for arbitrary
collections of polarizable neutral molecules \cite{Various}.  
For simplicity it is assumed
that all molecules have the same polarizability.  The generalization to
multiple species is straightforward.
The structure $\mu \approx f\alpha F_{\rm ext}$ is evident.
In sec. 4 it will be shown how the usual textbook CML follows from
this formula in the appropriate limit.

It is interesting and useful to consider the eigenvalues $\gamma$ and
eigenvectors $|\gamma>$ of the dipole-dipole interaction $\Gamma$:
\begin{equation}
{\bf{\Gamma}}=\sum_{\gamma}\gamma|\gamma><\gamma|. 
\label{eq:eigen}
\end{equation}
In terms of these eigenstates, the generalized CML for the induced dipoles is
\begin{equation}
\vec{\mu}_i=\sum_{\gamma}<i|\left(\frac{1}{\alpha}-\gamma\right)^{-1}
|\gamma><\gamma| F_{\rm ext}> 
\label{eq:cmleig}
\end{equation}
In the usual case, the factors $(1/\alpha-\gamma)^{-1}$
are all positive finite numbers.  The largest eigenvalue $\gamma$ is
less than the resistance $1/\alpha$ to polarization.  Because
the interaction Eq.(\ref{eq:gamma}) scales as $1/R^3$, the eigenvalues
$\gamma$ scale linearly with density $n$.  If one imagines
increasing the density of the system, leaving everything else fixed,
eventually the largest eigenvalue $\gamma_{\rm max}$ exceeds
$1/\alpha$ and the response is divergent.  This signifies that a
spontaneous polarization now occurs, with a pattern of 
polarization given by the corresponding eigenvector $|\gamma_{\rm max}>$.

\section{\label{sec:oct}Octahedral Fragments}

\begin{figure}[t]
\centerline{\psfig{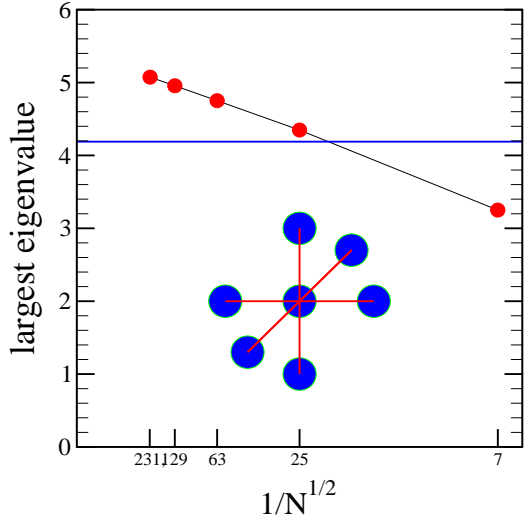}}
\caption{Largest eigenvalue of dipole-dipole matrix for octahedral fragments 
of a simple cubic lattice.  Units are used where the density $n$=1.  
The smallest octahedral fragment with 7 molecules is shown.  The
horizontal line is the CM eigenvalue $4\pi/3$.}
\label{fig:octeig}
\end{figure}

How does the CMM describes the evolution from nanocrystal
to bulk ferroelectric?  Consider the sequence of octahedral
fragments of a simple cubic lattice.  The simplest fragment with a central
molecule and six octahedral neighbors is shown in Fig. \ref{fig:octeig}.
Larger fragments have $N$=25, 63, ... molecules and retain the full cubic
point symmetry.  The largest eigenvalues of the dipole-dipole matrix have
been computed numerically for these cases and are shown plotted versus
a convenient reciprocal power of $N$ in Fig. \ref{fig:octeig}.
The $N=\infty$ bulk limit is nearly achieved for 200-molecule clusters.
For the infinite simple cubic crystal, the CML tells us that a ferroelectric
instability occurs when $1/\alpha$ diminishes to $4\pi/3$.  For fragments
of the crystal, the critical value of $1/\alpha$ should surely be less
than $4\pi/3$, increasing to this value as $N\rightarrow\infty$.  
Fig. \ref{fig:octeig} shows that there is a monotonic increase with $N$ but that
already by $N$=25 the largest eigenvalue exceeds the bulk CML value $4\pi/3$.
\par
\par

\begin{figure}[t]
\centerline{\psfig{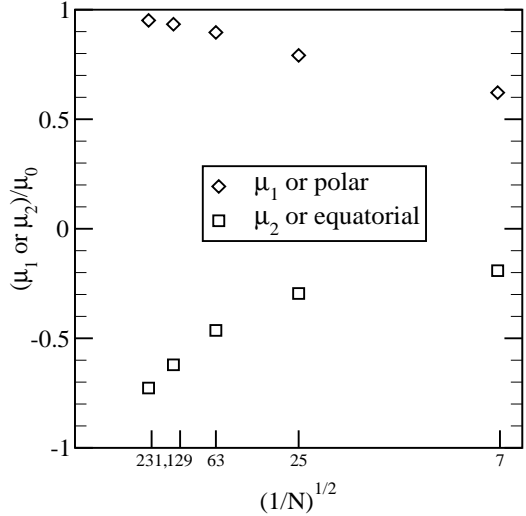}}
\caption{Eigenvector properties corresponding to the largest eigenvalue of 
the dipole-dipole matrix for octahedral fragments
of a simple cubic lattice.  The ``polar'' sites (immediately above
and below the central $z$-polarized site with polarization $\mu=1$)
have their polarization parallel and evolving toward 1 at large $N$.
The ``equatorial'' sites (four sites surrounding the central site in
the $xy$ plane) have their polarization antiparallel and evolving
toward -1.}
\label{fig:octevec}
\end{figure}

The reason for this unexpected result is that the {\it sc}
structure prefers antiferro rather than ferro ordering when
only dipole-dipole interactions are present.  
This was discovered by Luttinger and Tisza \cite{Luttinger}.
The next section 
elaborates, and Sec. \ref{sec:par} gives a simple way to calculate the 
preferred dipole ordering on a lattice.
To confirm the antiferro preference, Fig.\ref{fig:octevec} shows
how the polarization directions evolve as the octahedral fragment
enlarges toward the bulk limit.

\par
\par

A recent report by Fu and Bellaiche \cite{Fu} considers dipoles on
larger cubic fragments of a simple cubic lattice.  They find a
surprising spiral pattern to be stable, with no net moment.
They are solving a somewhat different model.

\section{\label{sec:bulk}Bulk Clausius-Mossotti Law}

Infinite periodic crystals have translational symmetry
that permits computation of the eigenvalues and eigenvectors
of the dipole-dipole matrix $\Gamma$.  Specifically, the 
Floquet theorem says that eigenstates $|\gamma>$
can be chosen to have the Bloch form $|\vec{k}\lambda>$ 
\begin{equation}
<i\alpha|\vec{k}\lambda>=e^{i\vec{k}\cdot\vec{R}_i} 
\epsilon_{\alpha}(\vec{k}\lambda)/\sqrt{N},
\label{eq:bloch}
\end{equation}
where (for a one-atom cell) the eigenvector 
$\hat{\epsilon}(\vec{k}\lambda)$
is a 3-d unit vector, labeled by wavenumber $\vec{k}$ and
branch index $\lambda$.  This eigenvector corresponds to eigenvalue
$\gamma(\vec{k}\lambda)$ and satisfies 
\begin{equation}
\Gamma_{\alpha\beta}(\vec{k})\epsilon_{\beta}(\vec{k}\lambda)
=\gamma(\vec{k}\lambda) \epsilon_{\alpha}(\vec{k}\lambda),
\label{eq:blocheig}
\end{equation}
where the $3 \times 3$ dipole-dipole matrix $\Gamma$ is now in Fourier space,
\begin{equation}
\Gamma_{\alpha\beta}(\vec{k})=\frac{1}{N}\sum_{i \neq j}
e^{i\vec{k}\cdot(\vec{R}_i -\vec{R}_j)}
\left( \nabla_{\alpha}\nabla_{\beta}\frac{1}{r} 
\right)_{\vec{r}=\vec{R}_i-\vec{R}_j}.
\label{eq:gammak}
\end{equation}

\par
\begin{figure}[t]
\centerline{\psfig{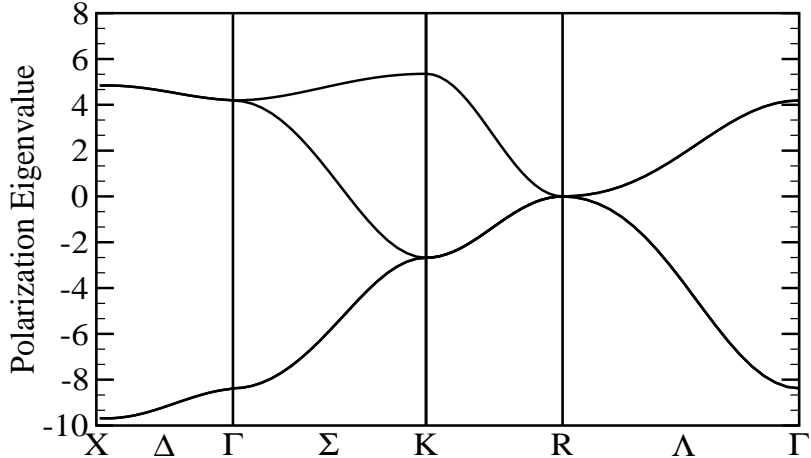}}
\caption{Wavevector dependence of the eigenvalues $\gamma(\vec{k}\lambda)$
of the dipole-dipole operator computed for the simple cubic structure,
in units where $n=1/a^3=1$.
Note that the largest eigenvalue is at $K=(\pi,\pi,0)$.
At $\Gamma=(0,0,0)$, the eigenvalues are $4\pi/3$ (the Clausius-Mossotti
value, doubly degenerate) and $-8\pi/3$ (singly degenerate).}
\label{fig:eigk}
\end{figure}
\par

For the {\it sc} structure, the eigenvalues $\gamma(\vec{k}\lambda)$
are shown in Fig.\ref{fig:eigk}.  The numerical calculation
used the Ewald method to converge the sums \cite{Smith}.
The sum of the three eigenvalues (trace
of the matrix $\Gamma_{\alpha\beta}(\vec{k})$)  
is 0, independent of $\vec{k}$.
This is because $\nabla^2(1/r)$ is 0 everywhere except at $r=0$
and the $\vec{R}_i=\vec{R}_j$ term is omitted from Eq.(\ref{eq:gammak}).
The same dispersion curve shown in Fig.\ref{fig:eigk}
has a different interpretation in phonon physics.  By a change
the sign and an additive constant $4\pi/3$, these become the
curves of squared frequencies of vibration of the
Coulomb lattice of unit point charges.  These are the
``bare phonons'' encountered in older approaches to the
theory of phonons in metals \cite{Pines}.  The fact that 
the eigenvalue at the $\Gamma$ point ($\vec{k}=0$) is not
the extreme eigenvalue in {\it sc} structure corresponds to a linear
instability (imaginary eigenfrequencies)
of the Coulomb lattice in {\it sc} structure.  
Conversely, the fact that the $\vec{k}=0$ eigenvalue $\gamma$ is maximal in
{\it fcc} and {\it bcc} structures corresponds to
linear stability of the Coulomb lattice in these structures.

The Clausius-Mossotti formula ($\mu_i=f\alpha \vec{F}_{\rm ext}$ with
$1/f=1-4\pi n \alpha/3$) can be obtained from these results. 
The external field is homogeneous, so only eigenvectors $|\gamma>$
corresponding to $\vec{k}=0$ can couple in Eq.(\ref{eq:cmleig}).
Then from Eq.(\ref{eq:cmleig}) we get 
\begin{equation}
\mu_i=\sum_{\lambda=1}^{3} \left(\frac{1}{\alpha} - 
\gamma(\vec{k}=0,\lambda) \right)^{-1}
(\hat{\epsilon}(\vec{k}=0,\lambda)\cdot\vec{F}_{\rm ext})
\hat{\epsilon}(\vec{k}=0,\lambda).
\label{eq:mucub}
\end{equation}
Thus it is needed that all three eigenvalues $\gamma(\vec{k}=0,\lambda)$
should be $4\pi n/3$.  Two eigenvalues indeed agree with this, but
consistent with summing to zero, the third eigenvalue is $-8\pi n/3$.
The corresponding eigenvector $\hat{\epsilon}(\vec{k}=0,\lambda)$ 
is ``longitudinal,'' that is, it pointed along $\vec{k}$ before the 
$\vec{k}\rightarrow 0$ limit was taken.  Longitudinal fields suffer
depolarization which transverse fields avoid.
A longitudinal field is created
by a parallel plate capacitor applied to a block of dielectric.
The ``external'' field is related to the classical macroscopic field
$\vec{F}$ by $\vec{F}_{\rm ext}=\vec{F}+4\pi \vec{P}$ and the polarization
density $\vec{P}$ is $n\vec{\mu}$.  
Thus for a longitudinal field, Eq.(\ref{eq:mucub}) is the same as
\begin{equation}
\vec{\mu}=\left(\frac{1}{\alpha} + \frac{8\pi n}{3}\right)^{-1}
(\vec{F} + 4\pi n \vec{\mu}).
\label{eq:long}
\end{equation}
This is equivalent to $\vec{\mu}=f\alpha\vec{F}$, which is the
correct Clausius-Mossotti law for a macroscopic sample.  

\section{\label{sec:fix}Fixed rather than Induced Moments}

The CMM refers to a system of polarizable molecules whose
induced dipoles $\vec{\mu}_i$ minimize the energy Eq.(\ref{eq:Emat}).  If 
spontaneous moments are just barely stable in no applied field, the  
solution is $\mu_{i\alpha}\propto<i\alpha|\gamma_{\rm max}>$,
where $|\gamma_{\rm max}>$ is the eigenvector of $\Gamma$ with
largest eigenvalue.
The proportionality constant is determined by higher order
terms.  

Another model, more in touch with chemical reality,
has molecules with moments $\vec{\mu}_i$ fixed in magnitude 
($|\vec{\mu}_i|=\mu_0$) but
free to choose their spatial orientation.  The ground-state
moment configuration then minimizes Eq.(\ref{eq:Emat}), except with
the resistance $1/\alpha$ to polarization set to zero, and with 
the $N$ constraints $|\vec{\mu}_i|=\mu_0$.  In general, this is
a difficult algebraic problem.  However, suppose the
eigenvector $|\gamma_{\rm max}>$ has the property that $|\vec{\mu}_i|=\mu_0$
where $\mu_{i\alpha}\propto<i\alpha|\gamma_{\rm max}>$.  For example,
an infinite system with perfect ferroelectricity or simple 
antiferroelectricity has this property (each moment has the same
magnitude.)  Then this also must provide the ground state of the fixed
moment problem.  As a magnetic example
(magnetic and electric fixed moments obey the same dipole-dipole laws),
White {\it et al.} \cite{White} studied
crystals of hydrated rare-earth phosphomolybdates.
The magnetic moments of the rare-earth ions (located
on a diamond-structure sublattice) are
fixed by Hund's rules but sufficiently separated from each
other that interatomic exchange is ignorable.  Then the
magnetic order (seen below an ordering temperatures $\le 50$ mK) is
determined by dipole-dipole interactions, and has been
measured.  The magnetic order is antiferromagnetic, much like
the result of Sec. \ref{sec:oct} for {\it sc} structure,
except the order is more interesting in diamond structure. 
As explanation, the authors computed the largest antiferromagnetic
eigenvalue of the matrix $\Gamma$ and showed that for diamond
structure (and also for {\it sc} but not for {\it bcc} or {\it fcc})
this eigenvalue exceeds the ferromagnetic (CM) eigenvalue $4\pi n/3$.
Their calculation is the $N\rightarrow\infty$ limit of Fig.(\ref{fig:octeig}).

In general, as illustrated in Fig. \ref{fig:octevec}, but 
unlike the bulk examples just discussed, 
the vector $|\gamma_{\rm max}>$ with largest eigenvalue 
of $\Gamma$ does {\bf not} have all moments 
$\mu_{i\alpha}\propto<i\alpha|\gamma_{\rm max}>$
equal in magnitude, and thus does not solve the problem of finding
the most favorable alignment of fixed dipole moments.
In what sense is the solution of the fixed dipole 
problem $\{\vec{\mu}_i\}$ ``close'' to the unconstrained 
solution given by $|\gamma_{\rm max}>$?  
It is required to minimize the dipole energy
$-<\mu|\Gamma|\mu>/2$ plus constraint terms $\lambda_i \mu_i^2 /2$.
The solution depends on the Lagrange multipliers (LM) $\lambda_i$ 
which are then varied until the constaints $\mu_i^2={\rm const}$ are
satisfied.  There are $N$ moments $\vec{\mu}_i$, but only $N-1$ 
constraints, because the constant value of $\mu_i^2$ is arbitrary 
since the problem is linear.  The minimization can be
formulated as a ``generalized Hermitean eigenvalue problem.''
The $N$ LM's define a vector $\vec{\lambda}$ in the $N$-dimensional
LM space.  Write this as $\vec{\lambda}=\lambda\hat{c}$ where
the components $c_i$ of $\hat{c}$ are normalized by $\sum c_i^2 = N$.
Also write $\mu_i^2=<\mu|P_i|\mu>$ where $P_i$ is a projection
operator onto the 3-vector space of the $i$-th moment $\vec{\mu}_i$.
Then the formal solution of the minimization problem is
\begin{equation}
\Gamma|\mu>-\lambda P(\hat{c})|\mu> =0
\label{eq:lagrange}
\end{equation}
where the Hermitean operator $P(\hat{c})\equiv\sum c_i P_i$
is the sum of projection operators.  The normalized vector $\hat{c}$
defines a ``ray'' in LM space.  There is one special ray $\hat{c}_0$, where
$c_i=1$ for all $i$, so that $P(\hat{c}_0)$ is the unit operator.  For this 
special case, Eq.(\ref{eq:lagrange}) is the ordinary eigenvalue
equation for $\Gamma$.  For other rays, as long as no component
$c_i$ is zero, $P$ is invertible and Eq.(\ref{eq:lagrange}) is
a generalized Hermitean eigenvalue equation.  The $N$ eigenvalues
$\lambda_{\kappa}$ depend on the direction $\hat{c}$ of the ray,
and define $N$ surfaces $\vec{\lambda}=\lambda_{\kappa}(\hat{c})\hat{c}$
in LM space.  Each surface $\kappa$ intersects the special ray at distance
$\gamma_{\kappa}$, where $\gamma_{\kappa}$ is an ordinary
eigenvalue of $\Gamma$.

All that has been accomplished so far is that the eigenvalue problem
for $\Gamma$ has been generalized to an infinite family of eigenvalue
problems, one for each ray $\hat{c}$ in LM space.  Now we have to
find the optimal solution $|\mu>$ which minimizes dipole energy
with fixed components $\mu_i=\mu$.  
The $N-1$ constraints $\mu_i^2={\rm const}$
will be satisfied along 1-dimensional lines in LM space.  These
lines will puncture each surface $\kappa$ at isolated points.
The optimal solution occurs at such a point on the surface
of maximum eigenvalues.  That is, it lies on the surface which
passes through $\gamma_{\rm max}\hat{c}_0$, {\it i.e.}, the surface
which evolves from the solution of the unconstrained problem.  
This suggests an algorithm.  First find the optimal
solution $|\gamma_{\rm max}>$ of the unconstrained problem.
Then find the nearby eigenvalue $\lambda_{\rm max}\approx\gamma_{\rm max}$
and corresponding eigenvector $|\lambda_{\rm max}>$
of the generalized problem for rays $\hat{c}$ near the 
special ray $\hat{c}_0$.  Wander stepwise through
nearby rays guided by the demand that the components
$\vec{\mu}_i =P_i |\lambda_{\rm max}>$ 
of the maximal eigenvector should evolve toward equal lengths.
This should provide an efficient iterative solution of the 
fixed moment problem.

\section{\label{sec:1d}1-Dimensional Dipole Stack}

Polar molecules in solution \cite{Teixeira} may tend to self-organize
into structures which optimize dipole-dipole interactions.
The ``Stockmayer fluid'' \cite{Leeuven} is a simplified model
for such a system, which keeps hard core repulsion and
long-range dipole-dipole interactions for spherical 
molecules with point dipoles.  Such models may show
linear arrangements of dipoles \cite{Gennes}.  Such arrangements
have also been directly visualized for magnetic nanoparticles
in solution \cite{Butter}, and even for 
Au nanoparticles \cite{Liao} which are presumed to acquire
their polarity from interactions with species
on their surface or in solution.  This section
considers ordered one-dimensional arrays, and the subsequent
two sections considers the interactions between different
ordered linear arrays. 

Consider a one-dimensional stack of dipoles,
with repeat distance $b$.
For simplicity, think of them as permanent rather
than induced moments, with identical magnitudes.
Let each moment $\vec{\mu}$ point along the
stack axis, chosen as the $z$ axis.
The total interaction energy is
\begin{eqnarray}
U_{\rm tot}&=&\frac{\mu^2}{2}\sum_{i\ne j}^N \frac{-2}{|z_{ij}|^3} 
\nonumber \\
&=& \frac{-2\mu^2}{b^3}\left[ \frac{N-1}{1^3}+\frac{N-2}{2^3}
+\frac{N-3}{3^3} + \cdots +\frac{1}{(N-1)^3}\right] 
\nonumber \\
&=& \frac{-2\mu^2}{b^3}\left[N\zeta(3)-\zeta(2)+\frac{1}{N-1/2}+\cdots\right]
\nonumber 
\label{eq:stacken}
\end{eqnarray}
Thus adding a molecule to the stack gains energy $2\zeta(3)\mu^2/b^3$
where $\zeta$ is the Riemann $\zeta$-function and $\zeta(3)$ has the
numerical value 1.202057$\ldots$.  Breaking a stack into two smaller
stacks creates two new surface points, each costing
energy $(\mu^2 /b^3)\zeta(2)$ where $\zeta(2)=\pi^2 /6 = 1.6449\ldots$.

If the molecules are instead polarizable (with zero fixed moment) then
the stack will spontaneously polarize when the energy gain exceeds
the cost $\mu^2/2\alpha$, that is, when $1/\alpha$ is less than
$4\zeta(3)/b^3$.  One should ask if this situation is thermodynamically
stable.  For example, how much does it cost to rotate the
direction of $\vec{\mu}$ away from the stack axis?  How much does it
cost to create a defect, for example, a stacking fault where the
direction of $\vec{\mu}$ changes at the fault?

For the octahedral fragments considered in Sec. \ref{sec:oct}, cubic point
symmetry has the consequence that until terms of higher order are
considered, there is no preference for any particular direction
of $\vec{\mu}$.  There is no anisotropy energy, so the ordered
moment lacks a permanent direction in space.  A collection of such
nanoparticles would be called a ``superparaelectric,'' the analog of
a ``superparamagnet.''  For the 1-d stack,
this is no longer the case.  Cubic symmetry is maximally broken,
and it costs $3\zeta(3)\sin^2(\theta)\mu^2/b^3$ per molecule
to rotate all dipoles.  The anisotropy energy is very large.

On the other hand, it is a well-known property of ordered states
in one dimension that domain wall defects destroy long-range order.
We can estimate the cost of a domain wall as follows.  An
absolutely abrupt boundary, where induced moments have their
full positive value on one side of the boundary between two
molecules, and their full negative value on the other side, costs
energy $4\zeta(2)\mu^2/b^3$ in lost dipole-dipole
attraction.  A more gradual case is an abrupt boundary on
the site of a molecule.  This loses the same amount of
dipole-dipole attraction but there is a savings $\mu^2/2\alpha$
since one molecule is unpolarized, making this domain wall structure
more favorable.  More gradual boundaries
become lower in energy when the polarizability is diminished.

Domain-wall defects will always be thermally excited except in shorter
chains at lower temperature.  The number of thermally excited
domain walls is $Np$ where $p=\exp(-\Delta/k_B T)$ and $\Delta$ is the
energy to create a domain wall.  Suppose molecules in
solution interact by dipolar forces.  It is argued in Sec. \ref{sec:ell}
that for oblate ellipsoidal molecules with dipoles along
the short axis, chains are favored and only weakly attract
neighboring chains.  Then the statistical distribution of
chain lengths $n$ may be given by the Flory-Schulz \cite{Peebles}
distribution $P_n=p(1-p)^{n-1}$.  The mean chain length
would be $1/p$ but the most probable chain length would be 1.
The derivation of this distribution assumes that $\Delta$ and
thus $p$ do not depend on length, whereas actually $\Delta$ 
becomes smaller for short chains, which will enhance $P_n$ at
small $n$.  The Flory-Schulz distribution is usually
written in terms of a variable $p^{\prime}=1-p$ which
measures the ``completion'' of the ``reaction.''
For prolate ellipsoids, there will be greater organization of
chains into bundles.  Domain walls then become correspondingly 
larger in area and
harder to excite thermally, and the system evolves into a 
ferroelectric.

\section{\label{sec:par}Interacting Parallel Stacks}

There is an interesting and important interaction between
two parallel stacks of fixed dipoles.  Consider the
case of infinite stacks (vertical separation of moments
is $b$), where each moment is vertical (along the stack) and the
two stacks are identical and parallel.  Then it is obvious
that the interaction is a periodic function of the vertical
offset $y$ with period $b$.  It turns out that
the interaction is strictly oscillatory with no constant
component (in powers of $\exp(i2\pi y/b)$) and 
decays exponentially (as $\exp(-2\pi x/b)$)
with horizontal separation $x$.  

The stack-stack interaction (per molecule) is
\begin{equation}
U_{\rm ss}=\frac{\mu^2}{b^3}\sum_{n=-\infty}^{\infty}
\frac{1}{[x^2+(nb-y)^2]^{3/2}}\left[ 1 - \frac{3(nb-y)^2}
{x^2+(nb-y)^2} \right].
\label{eq:ssint}
\end{equation}
This is also the interaction energy between a single infinite stack
and a single $z$-oriented dipole at $(x,y)$.
At large separations $x$ we can use the continuum approximation
of a constant linear dipole density $\mu/b$ along the stack.  
Then the stack is electrostatically equivalent to 
charges $\pm \mu/b$ on the ends of the stacks which are infinitely
removed from the observation point.  Therefore at large separations
the electric field and the interaction are zero.  
Any non-zero remainder (contained
in Eq.(\ref{eq:ssint}) derives from the graininess of the dipole
density.

\begin{figure}[t]
]\centerline{\psfig{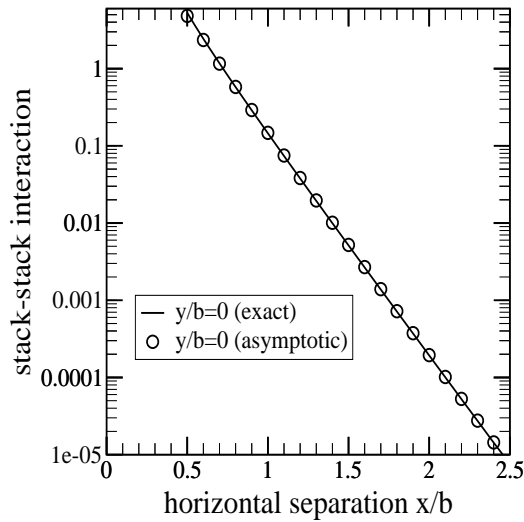}}
\caption{Stack-stack interaction between two identical stacks of
fixed dipoles $\mu$ separated along the stack by distance $b$.  The
stacks are separated horizontally by distance $x$ shown horizontally
in units of $b$.  The vertical offset $y$ is zero.  The interaction is
in units $\mu^2/b^3$.  The solid line is computed numerically from
Eq.(\ref{eq:ssint}) and the circles are the first term, 
Eq.(\ref{eq:asymint}) in the asymptotic expansion.}
\label{fig:logint}
\end{figure}

\begin{figure}[t]
\centerline{\psfig{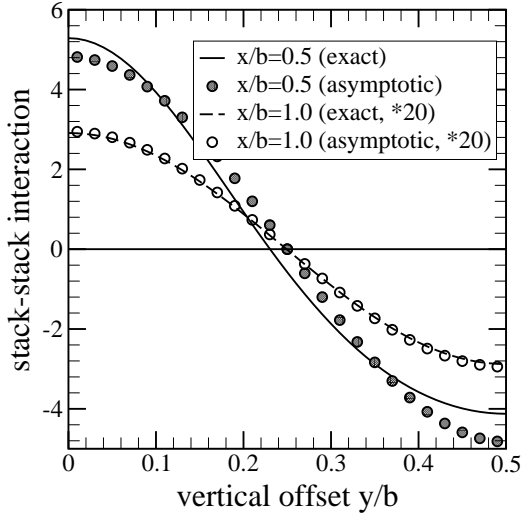}}
\caption{Stack-stack interaction between two identical stacks of
fixed dipoles $\mu$ separated along the stack by distance $b$.  The
stacks are separated horizontally by distance $x/b=0.5$ 
(solid line and filled circles) or by distance $x/b=1.0$ (dashed line
and open circles.  The vertical offset $y/b$ is shown horizontally.
The interaction is in units $\mu^2/b^3$.  Note that the $x=b$
interaction is shown after multiplication by 20.}
\label{fig:inty}
\end{figure}

To derive an asymptotic formula for the interaction,
use the Poisson sum formula
\begin{equation}
\sum_{n=-\infty}^{\infty}f(in)=\frac{1}{2i}\int_C dz \coth(\pi z)f(z).
\label{eq:poisson}
\end{equation}
The contour in the complex $z$-plane surrounds the imaginary
axis where $coth(\pi z)$ has poles at $z=in$.  Rewrite
the summand of Eq.(\ref{eq:ssint}) as 
$-d^2/dy^2[x^2+(nb-y)^2]^{-1/2}$.  After bending the contour
around suitable branch cuts, the dimensionless interaction is
\begin{equation}
\frac{U_{\rm ss}}{\mu^2/b^3}=u_{\rm ss}\left(\frac{x}{b},\frac{y}{b}\right)=
-2\frac{d^2}{d(y/b)^2}
\int_{x/b}^{\infty}\frac{du}{\sqrt{u^2-(x/b)^2}}
\frac{\sinh(2\pi u)}{\cosh(2\pi u) - \cos(2\pi y/b)}
\label{eq:ssint1}
\end{equation}
From this exact representation, extract the leading
term in the asymptotic expansion 
in powers $\exp(2\pi n x/b)$ and $(x/b)^{-m-1/2}$,
\begin{equation}
u_{\rm ss}\left(\frac{x}{b},\frac{y}{b}\right)
\approx 8\pi^2 \cos(2\pi y/b)  \frac{e^{-2\pi x/b}}{\sqrt{x/b}}.
\label{eq:asymint}
\end{equation}
An understanding of formulas like this dates back to Madelung
\cite{Madelung}.  Because of $y$-periodicity, the electrostatic
potential $\phi(x,y)$ can be expanded as $\sum C_{\ell}f_{\ell}(x)
\exp(2\pi i \ell y/b)$.  Madelung points out that Poisson's equation
demands that $f_{\ell}(x)$ be the Hankel function $K_0(2\pi \ell x/b)$.
The absence of the constant ($\ell=0$) term from the sum follows
from the charge neutrality of the stack.  The asymptotic behavior
of $K_0(t)$ is $\exp(-t)\sqrt{\pi/2t}$.
In Figs.\ref{fig:logint} and \ref{fig:inty}, the exact and the asymptotic 
version of the interaction are compared.  They agree 
to within $\pm 15\%$ for $(x/b)=0.5$ and 
to within $\pm 2\%$ at $(x/b)=1$.  A plot of the electric field
near the stack is shown in a separate paper \cite{Allen}
\par
\par
\par
\par

The oscillation in $u_{ss}$ as a function of $y/b$ shown in 
Fig.\ref{fig:inty} has interesting consequences.
When adjacent stacks
have zero offset (as happens in the {\it sc} crystal structure),
their moments $\vec{\mu}$ prefer to be antiparallel, as was 
already seen for the octahedral fragments in Sec. \ref{sec:oct}.
When adjacent stacks have a full offset $y/b=0.5$, their
moments prefer to be parallel, which is known to be the preference
in the {\it bcc} and {\it fcc} crystal structures.  Triangular
packing of adjacent stacks has ``frustrated'' interactions.
If the offset is zero, then some nearest neighbor interactions
are necessarily repulsive no matter what orientations of $\vec{\mu}$
are chosen.  This problem is discussed further later in this section
and in the next. 

For cubic structures, an interesting sum-rule
on stack-stack interactions is implied by the
CM formula.  This formula is equivalent
to the statement that the dipole-dipole interaction
energy for a cubic arrangement of dipoles, all oriented in the
same direction, is $-\gamma_{\rm CM}\mu^2 /2$ per molecule, and the
CM eigenvalue $\gamma_{\rm CM}$ is $4\pi n/3$.
Crystal structures can be viewed (in multiple ways)
as parallel lines of stacked 
molecules.  If the dipoles line up along the stacks,
the total dipole-dipole energy can be computed
as the energy per molecule in a stack, plus the energy per molecule
of stack-stack interactions.  Set these two versions of the
total energy equal to each other, divide out $\mu^2$, and
multiply by the volume of a cube whose edge length is the 
periodicity $b$ on the stack.  Define $\nu=nb^3$.  Then 
the dimensionless energy equation is the sum rule
\begin{equation}
-\frac{2\pi\nu}{3}=-2\zeta(3) + \frac{1}{2}\sum_{j\neq 0}^{\rm stacks}
u_{\rm ss}\left(\frac{x_j}{b},\frac{y_j}{b}\right) \ \ \ 
{\rm (cubic \ structures)}
\label{eq:sumrule}
\end{equation}

Because the stack-stack interaction decays rapidly with separation,
the dipole-dipole energy comes from the single stack energy 
$-2\zeta(3)=-2.4041$
plus near neighbor stack interactions (the last term
in Eq. \ref{eq:sumrule}, with a factor of 1/2 to avoid double counting.)
This is demonstrated to be true in tables 1--3.  For each
crystal structure, there is more than one simple axis which can be
chosen as a stack axis.  In {\it bcc} structure 
(111)-oriented stacks are a favorable choice relative to (100) 
orientation because the spacing of molecules is closer ($b=\sqrt{3}a/2$
instead of $a$);  (110) stacks are best
for both diamond and {\it fcc} structures, while
(100) stacks are best in {\it sc} structure.  The sum rule applies
for any consistent choice of stacks.  First-neighbor interactions 
converge the sum rule to better than 1\% for {\it sc} (100) and 
{\it bcc} (111) stacks, and {\it fcc} requires only second-neighbor
interactions in either (100) or (110) orientations.  It is easily
seen how the {\it sc} structure lowers its dipole energy by
switching orientation of first-neighbor (100) stacks.  For diamond,
although (110) stacks give only a small improvement in convergence, 
they allow a better understanding of the antiferroelectric solution
which gives the largest eigenvalue $\gamma$.  Each (110) stack has one
very close neighbor with offset $y/b=1/2$ which makes a strong
ferroelectric bond, and two quite close next neighbors with zero offset,
which therefore prefer to be antiferroelectrically oriented.  

Solutions of polar molecules of ellipsoidal shape may self-organize
into arrays of stacks.  For cylindrically shaped molecules, the
stacks may organize in a 2-d triangular lattice.
Therefore it is interesting to look
at the cubic structures from this point of view.  Diamond is the only
cubic structure that does not work,
because (111) stacks are dimerized (alternate spacings
of $\sqrt{3}a/4$ and $3\sqrt{3}a/4$ occur). 
The other cubic structures are triangular arrays of (111) stacks.  
The ``frustration'' problem is solved by alternating offsets of $\pm 1/3$
for the six first-neighbor stacks surrounding a central stack.
At this offset, ferroelectric alignment is preferred by all first-neighbor
pairs.  Consider the rhombohedral translation vectors $\vec{R}_i=
\vec{R}(\theta_i)=r\cos\theta_i \hat{x} +r\sin\theta_i \hat{y} +(b/3)\hat{z}$, 
where $\theta_i=0^{\circ}$, $120^{\circ}$, and $240^{\circ}$.  
The sum of these three primitive translations is $b\hat{z}$, the period
of the stack, and the distance $r$ is the transverse separation of stacks.
The rhombohedral angle $\alpha$ is defined by
$R^2 \cos\alpha = \vec{R}_i \cdot \vec{R}_j$ which gives
$\cos\alpha=[(b/3)^2-r^2 /2]/[(b/3)^2+r^2]$.  The three special values
of $\cos\alpha$ are -1/3 ({\it bcc}), 0 ({\it sc}), and 1/2 ({\it fcc}),
which correspond to ratios $r/b$ of spacing to period of $r_1=\sqrt{8/9}$
({\it bcc}), $r_1 /2$ ({\it sc}), and $r_1 /4$ ({\it fcc}).
For {\it bcc}, the
period and separation are nearly the same, so stacks
interact only weakly.  The {\it sc}
structure is intermediate, while {\it fcc} structure has
a small transverse separation relative to periodicity, leading to very
strong stack-stack interactions.  
For both {\it fcc} and {\it bcc} structures, the ferroelectric
alignment is optimal.  However, for the intermediate case of {\it sc}
structure, a lower energy solution tilts the axis of 
polarization of molecules along (111) stacks alternately to $\pm(100)$.
Structures like {\it sc} and {\it fcc} should not necessarily be
considered as arrays of (111) stacks even though it may be 
mathematically interesting to do so.

\vskip 4mm
\begin{center}
\begin{tabular}{|r|rrrrrr|}   \hline\hline
structure & $\nu$ & $\frac{-2\pi\nu}{3}+2\zeta(3)$ & $x_j$ & $y_j$ & $M_j$ & 
        $\frac{1}{2}M_j u_{\rm ss}(x_j,y_j)$\\ \hline
{\bf sc}  & 1 &  0.3097 & 1                    & 0             & 4 &  0.2910 \\ 
    \hline
{\bf bcc} & 2 & -1.7847 & $\frac{1}{\sqrt{2}}$ & $\frac{1}{2}$ & 4 & -2.0822 \\
          &   &         & 1                    & 0             & 4 &  0.2910 \\ 
    \cline{7-7}
          &   &         &                      &               &   &  
           $\sum=$  -1.7912 \\ \hline
{\bf fcc} & 4 & -5.9735 & $\frac{1}{2}$        & $\frac{1}{2}$ & 4 & -8.2484 \\
          &   &         & $\frac{1}{\sqrt{2}}$ & 0             & 4 &  2.2272 \\
    \cline{7-7}
          &   &         &                      &               &   & 
           $\sum=$  -6.0212 \\ \hline 
{\bf diamond}&8&-14.3510& $\frac{1}{2\sqrt{2}}$& $\frac{1}{4}$ & 4 & -8.3287 \\
          &   &         & $\frac{1}{2}$        & $\frac{1}{2}$ & 4 & -8.2484 \\
          &   &         & $\frac{1}{\sqrt{2}}$ & 0             & 4 &  2.2272 \\
          &   &         & $\frac{\sqrt{10}}{4}$& $\frac{1}{4}$ & 8 & -0.0481 \\
          &   &         & 1                    & 0             & 4 &  0.2910 \\
          &   &         & $\frac{3}{2\sqrt{2}}$& $\frac{1}{4}$ & 4 & -0.0007 \\
          &   &         & $\frac{\sqrt{5}}{2}$ & $\frac{1}{2}$ & 8 & -0.2606 \\ 
    \cline{7-7}
          &   &         &                      &               &   & 
           $\sum=$  -14.3683 \\ \hline 
\hline
\end{tabular}
\\
{\bf Table. 1} Sum rule satisfaction for cubic crystals where the
stacks have period $b=a$ along a cube axis, and dipoles are oriented
along the same axis.  $\nu$ is the number of molecules in a cube of 
edge $b$.  $M_j$ is the multiplicity of the $j$-th set of neighboring stacks.
\end{center}
\vskip 4mm
\vskip 4mm
\begin{center}
\begin{tabular}{|r|rrrrrr|}   \hline\hline
structure & $\nu$ & $\frac{-2\pi\nu}{3}+2\zeta(3)$ & $x_j$ & $y_j$ & $M_j$ &
        $\frac{1}{2}M_j u_{\rm ss}(x_j,y_j)$\\ \hline
{\bf fcc} & $\sqrt{2}$ &-0.5578 & $\frac{\sqrt{3}}{2}$ & $\frac{1}{2}$ & 4 
          & -0.7110 \\
          &   &         & 1    & 0             & 2 &  0.1455 \\
    \cline{7-7}
          &   &         &                      &               &   &
           $\sum=$  -0.5655 \\ \hline 
{\bf diamond}&$2\sqrt{2}$&-3.5196& $\frac{1}{2\sqrt{2}}$& $\frac{1}{2}$ & 1 
          & -5.0847 \\
          &   &         & $\frac{\sqrt{6}}{4}$ & 0 & 2 & 2.2235 \\
          &   &         & $\frac{\sqrt{3}}{2}$ & $\frac{1}{2}$ & 4 & -0.7710 \\
          &   &         & 1                    & 0 & 2 & 0.1455 \\
          &   &         & $\frac{3\sqrt{2}}{4}$& $\frac{1}{2}$ & 3 & -0.1436 \\
          &   &         & $\frac{\sqrt{22}}{4}$& 0             & 2 &  0.0454 \\
    \cline{7-7}
          &   &         &                      &               &   &
           $\sum=$  -3.5249 \\ \hline
\hline
\end{tabular}
\\
{\bf Table. 2} Sum rule satisfaction for cubic crystals where the
stacks have period $b=a/\sqrt{2}$ along a cube face diagonal.
$\nu$ is the number of molecules in a cube of edge $b$.
\end{center}
\vskip 4mm
\vskip 4mm
\begin{center}
\begin{tabular}{|r|rrrrrr|}   \hline\hline
structure & $\nu$ & $\frac{-2\pi\nu}{3}+2\zeta(3)$ & $x_j$ & $y_j$ & $M_j$ &
        $\frac{1}{2}M_j u_{\rm ss}(x_j,y_j)$\\ \hline
{\bf bcc} & $\frac{3\sqrt{3}}{4}$ & -0.3166 & $\sqrt{8/9}$ 
          & $\frac{1}{3}$ & 6 & -0.3224 \\ \hline
{\bf sc}&$3\sqrt{3}$&-8.4787& $\sqrt{ 2/9}$ & $\frac{1}{3}$ & 6 & -9.6448 \\
        &           &       & $\sqrt{ 6/9}$ & 0             & 6 &  1.5423 \\
        &           &       & $\sqrt{ 8/9}$ & $\frac{1}{3}$ & 6 & -0.3224 \\
        &           &       & $\sqrt{14/9}$ & $\frac{1}{3}$ &12 & -0.0826 \\
    \cline{7-7}
          &   &         &                      &               &   &
           $\sum=$  -8.5075 \\ \hline
\hline
\end{tabular}
\\
{\bf Table. 3} Sum rule satisfaction for cubic crystals where the
stacks have period $b=\sqrt{3}a/2$ ({\it bcc}) and $b=\sqrt{3}a$ ({\it sc})
along a cube body diagonal.
$\nu$ is the number of molecules in a cube of edge $b$.  The sum rule is
satisfied to similar accuracy for {\it fcc} structure with (111) stacks
only when stacks out to 12th neighbors are summed.
\end{center}
\vskip 4mm

\section{\label{sec:ell}Close Packed Ellipsoidal Dipoles}

Here is a mathematical puzzle of possible chemical interest.
It is a relative of a popular statistical mechanics model,
the dipolar liquid \cite{Teixeira}.
Consider an infinite set of hard ellipsoids of revolution
(radius of revolution $a/2$, axial half-length $b/2$.
Let there be a permanent moment $\mu$ along the unique axis.
What is the ground state structure and energy?  
This section
only states and does not pretend to solve the problem,
but makes a conjecture about the answer for the special
case $a=b$ of hard-sphere dipoles.
This puzzle would be relevant to liquid phase self-assembly of 
ellipsoidal shaped molecules, if
no other interactions were important besides hard-core 
repulsion and electrostatic interactions. 

One possible family of structures is arrays of ``chains,'' where 
the term ``chain'' will be used to denote a stack of
touching ellipsoids with the axis of revolution 
the same as the axis of the chain.  Dipole-dipole interactions
will favor ferro alignment along the chain.  What transverse
arrangement of chains is favored?  Close packing will require
at least an approximate triangular pattern of chain axes.
In the ``simple hexagonal'' structure of perfect triangular
symmetry and zero longitudinal offsets, first neighbor
chains all wish to have antiparallel moments.  This is geometrically
impossible, so the pattern is frustrated. Various anti-ferroelectric
({\it e.g.} alternating rows of up and down oriented chains) or ferrielectric
({\it e.g.} one out of three anti-aligned chains in a hexagonal pattern) give
two out of three neighbors favorably anti-aligned and one out of three
unfavorably aligned.  

The ``offset'' is defined as the distance along the chain
axis separating centers of ellipsoids on two neighboring
chains, measured in units of the period $b$ which is also
the axial length of the ellipsoid.  With offset zero,
touching ellipsoids have transverse separation $r=a$.
With non-zero offset, the separation is $r=a\sqrt{1-({\rm offset})^2}$.
Introducing offsets allows two good things to
happen.  (1) The chains can pack more densely because the widest parts of
adjacent ellipsoids are not touching.  (2) For offsets roughly between
1/4 and 1/2, parallel moments are preferred.  If all neighbors prefer
to be parallel, there is no frustration problem.  Of course, farther
neighbors have other offsets, and if $b/a>1$, farther interactions
acquire more importance, reintroducing frustration.

\begin{figure}[t]
\centerline{\psfig{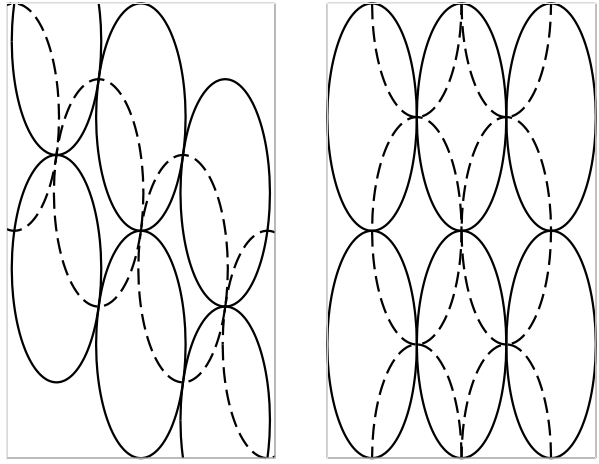}}
\caption{Rhombohedral (left) and body-centered orthorhombic ({\it bco}, right)
packing of hard ellipsoids.  The axis of revolution is shown vertical.
Rhombohedral packing achieves a greater 2-d density of ellipsoids, but
lesser 3-d density, than {\it bco}.}
\label{fig:ell}
\end{figure}

From the discussion of the previous
section, two symmetrical possibilities can be seen:  (1) a triangular 
array with offsets
$\pm 1/3$, which gives the rhombohedral family to which 
the primitive cubic structures belong; or (2) a body-centered
orthorhombic ({\it bco}) family to which the {\it fcc} structure
belongs.  These structures are shown in Fig. \ref{fig:ell}.
The {\it bco} structure minimizes the frustration problem by
breaking triangular symmetry, giving the four closest neighbors
the optimal offset 1/2 for ferroelectric order, and having
two slightly more distant neighbors with zero offset which are
unhappily ferroelectric.

It is easy to see that {\it bco} is the preferred
structure for dipolar hard
spheres.  Note that the density of ellipsoids 
($1/v$ where $v$, the volume per ellipsoid, is $a^2 b/\sqrt{2}$ 
for {\it bco} and $\sqrt{16/27}a^2 b$ for rhombohedral) is higher
for {\it bco} at all axial ratios $b/a$.  For cubic structures
(these structures are {\it fcc} and {\it bcc} respectively when
$b=a$) the CM law tells us that the dipole energy $E=-(2\pi/3v)\mu^2$
depends only on the density.  For other axial ratios, the CM
law no longer works because the structure is not cubic.  One 
might expect that {\it bco} would always win because of its
higher density.  However, numerical calculation shows that for
axial ratios $b/a > 1.717$, the rhombohedral structure is better.
This numerical test was done only in the range $0.5 < b/a < 2.5$.
For larger $b/a$, the on-chain parallel alignment of dipoles is 
no longer a big part of the energy, so there is no reason to think
that chains would form.  For smaller $b/a$, the chain-chain interaction
is no longer a big part of the energy, so it is not likely that 
dipole interactions would play the dominant role in determining
how chains arrange themselves.

\par
\par

It seems reasonable to conjecture that the optimal solution
for hard sphere dipoles is the {\it bco}={\it fcc}
structure with minimum volume.  However, the complexity of
dipolar interactions is such that it is not wise to be to
confident of this solution.  For ellipsoids that are close
to spherical, the same answer probably applies, but for
$b/a=1.717$ where {\it bco} loses to rhombohedral, there
is no good argument preventing a totally different structure
from doing even better.

\section{\label{sec:met}Metallic Coatings}

The CMM focusses on a single aspect of the physics of electrical
polarity in nanosystems, namely dipole-dipole interactions of induced
moments.  In this final section the model is pushed a little 
further to model the effect of metallic screening from adjacent
metallic regions.  An interesting result was derived by Maschhoff
and Cowin \cite{Maschhoff}.  They find that a neutral molecule
with both a fixed point dipole moment $\mu_0$ and an isotropic
polarizability $\alpha$, when favorably aligned perpendicular
to a metal surface at distance $b/2$, has energy (attraction
to the image dipole) 
\begin{equation}
E=-\mu_0^2/(b^3  -2\alpha).
\label{eq:metsurf}
\end{equation}
The induced dipole enhances the total dipole and increases
the attractive interaction energy between the dipole and the
surface.  At a critical polarizability $\alpha_c=b^3 /2$,
the response is divergent.  At this point, an induced dipole
would spontaneously form even if the permanent moment $\mu_0$
were zero.

It is evident that coupling a polarizable medium to 
metallic electrodes or shielding devices may promote
spontaneous polarization.  As a specific model, we now
investigate an array of $N$ fixed 
polarizable molecules (positions $\vec{R}_i$) which
are coated, or otherwise mingled, with $N^{\prime}$ fixed chargeable molecules 
(positions $\vec{R}_{\ell}$).
The indices for polar sites are $i,j$ and for charge sites are 
$\ell,m$.  Many models of this type can be found in
the literature, for example 
\cite{Dugourd,Rubio,Goddard,Friesner,Madden,Pederson}.
The energy of chargeable sites alone would be the
classical electrostatic energy
\begin{eqnarray}
{\cal E}_{\rm ch}(\{q_{\ell}\})&=&\frac{1}{2C}\sum_{\ell}^{N^{\prime}}
q_{\ell}^2 +\frac{1}{2}\sum_{\ell\neq m} \frac{q_{\ell}q_m}{R_{\ell m}}
+ \sum_{\ell} q_{\ell}\phi_{\ell,{\rm ext}}
\nonumber \\
&=& \frac{1}{2} <q|[(1/C)1 + V]|q>+<q|\phi_{\rm ext}>. 
\label{eq:elecen}
\end{eqnarray}
The term $q_{\ell}^2/2C$ is the energy cost of putting charge
$q_{\ell}$ at the $\ell$'th site.  Here $|q>$ is an $N^{\prime}$-vector
and $1$ and $V$ are $N^{\prime}\times N^{\prime}$ matrices.
The external potential $\phi_{\ell,{\rm ext}}=
\phi_{\rm ext}(\vec{R}_{\ell})$ is the external scalar potential,
related to the external electric field
by $\vec{F}_{\rm ext}=-\vec{\nabla}\phi_{\rm ext}(\vec{r})$.
The charges can adjust to minimize energy, subject to the constraint
of charge conservation, $\sum_{\ell} q_{\ell}=$const, where the constant
will normally be zero.  Introducing a Lagrange multiplier $\lambda$,
we find that for a system of charge sites in constant external potential,
\begin{equation}
[(1/C)1+V]|q>+|\phi_{\rm ext}>=\lambda|0>
\label{eq:elpot}
\end{equation}
where $<\ell|q>=q_{\ell}$ is the charge at site $\ell$ and 
$<\ell|0>\equiv 1$ defines $|0>$.  
The first term on the left hand side of Eq.(\ref{eq:elpot}) is the
vector $|\phi_{\rm int}>$ of internal electrostatic potentials at sites.
The charges adjust to make a constant potential $\phi=\phi_{\rm int}
+\phi_{\rm ext}$.
A non-zero $\lambda$ just uniformly shifts the potential
in order to adjust the total charge.  The system
of charges should be neutral when not driven.  This requires
the matrix $[(1/C)1+V]$ to be positive.  The Coulomb
interaction $V$ by itself is not positive -- two sites
with charge $+Q$ and $-Q$ have negative energy -- so a
capacitance $C$ less than infinity is needed to keep
charges from spontaneously separating.  In the resting state
($\phi_{\rm ext}=0)$
we want a neutral system ($<0|q>=0$) so $\lambda=0$.

Now introduce electrostatic coupling between the charge sites
and the polar sites,
\begin{eqnarray}
{\cal E}_{\rm int}(\{q_{\ell}\},\{\mu_i\})
&=&\sum_{\ell}^{\rm charges} \sum_{i}^{\rm dipoles}
\frac{q_{\ell}\vec{R}_{\ell i}\cdot\vec{\mu}_i} {R_{\ell i}^3}
\nonumber \\
&=& <q|K|\mu>
\label{eq:chdipint}
\end{eqnarray}
where the matrix K has dimension $N^{\prime}\times 3N$.
For simplicity, choose $\phi_{\rm ext}=0$, and
find the charge distribution $|q>$ which minimizes the total
energy, the sum of Eqs.(\ref{eq:Emat},\ref{eq:elecen},\ref{eq:chdipint}),
for an arbitrary fixed dipole distribution $|\mu>$.
The answer is
\begin{equation}
|q>=-\left(\frac{1}{C} +V\right)^{-1}K |\mu>.
\label{eq:chdist}
\end{equation}
When this answer is inserted into the total energy formula,
the result is an effective dipole-dipole interaction $\Gamma_{\rm eff}$,
\begin{equation}
{\cal E}_{\rm tot} = \frac{1}{2} <\mu|[(1/\alpha)1-\Gamma_{\rm eff}|\mu>
\label{eq:defgameff}
\end{equation}
\begin{equation}
\Gamma_{\rm eff}= \Gamma + K^{\dagger}[(1/C)1+V]^{-1} K
\label{eq:gameff}
\end{equation}
where $K^{\dagger}$ is the transpose of the rectangular matrix 
$K$.  The charge system effectively enhances the dipole-dipole
interaction, driving the system closer to a ferroelectric
instability.  

As an illustration, consider a single polarizable molecule,
with two chargeable sites, one to each side at a distance $r$.
Assuming that charge responds instantly to polarization
on the polarizable molecule, the energy to create a
dipole is $\mu^2/2\alpha_{\rm eff}$, where
\begin{equation}
\frac{1}{\alpha_{\rm eff}}=\frac{1}{\alpha}
-\frac{1/r^4}{1/2C-1/4r}.
\label{eq:alphaeff}
\end{equation}
The $1/r^4$ in the numerator is the two powers of the charge
dipole interaction $K$, and $1/4r$ in the denominator is half the
Coulomb attraction between the induced charges $\pm q$.
When there is an external field (along the trimer axis) 
the total dipole $\mu_{\rm tot}=\mu+2qr$ has both a contribution
from the polar molecule and from the metal, and the response is
\begin{equation}
\mu_{\rm tot}=\alpha_{\rm eff}\left[1+\frac{1/\alpha-2/r^3}
{\frac{1}{2r^2}\left(\frac{1}{C}-\frac{1}{2r}\right)}\right]F_{\rm ext}
\label{eq:resp}
\end{equation}
This diverges at the same critical polarizability that makes
$\alpha_{\rm eff}$ diverge.

It is easy to produce an alternative theory in which charges
are treated by quantum mechanics, and charge interactions
are solved in random phase approximation.  The simplest such
model is to let the charge sites $\vec{R}_{\ell}$ be atoms
which have only an $s$-orbital available near the Fermi level.
The Hamiltonian is
\begin{equation}
{\cal H}_{\rm el}=\sum_{\ell m} t_{\ell m}c^{\dagger}_{\ell}c^{\ }_m
+\frac{1}{2}\sum_{\ell \ne m} V_{\ell m}c^{\dagger}_{\ell}c^{\ }_{\ell}
  c^{\dagger}_m c^{\ }_m
\label{eq:Ham}
\end{equation}
where $c_m$ destroys an electron from the orbital on site $\vec{R}_m$.
The charge-dipole interaction is
\begin{equation}
{\cal H}_{\rm el-dip}=\sum_{i\alpha m}K_{m,i\alpha}
\mu_{i\alpha}c^{\dagger}_m c^{\ }_m
\label{eq:heldip}
\end{equation}
Now the induced charge at site $\ell$ is related to the external
potential at site $m$ by the density response function $\chi$,
\begin{equation}
\delta q_{\ell}=-\sum_m \chi_{\ell m} \delta V_{{\rm ext},m}.
\label{eq:susc}
\end{equation}
and the energy of the induced charge distribution
(analog of Eq.(\ref{eq:elecen}) is
\begin{equation}
{\cal E}_{\rm ch}(\{q_{\ell}\})=\frac{1}{2}<q|\chi^{-1}|q>
\label{eq:suscen}
\end{equation}
If we ignore the Coulomb interaction (second term in Eq.(\ref{eq:Ham})),
then the electrons are governed by the non-interacting Hamiltonian
${\cal H}_{\rm el}^0$.  This problem has the solutions
${\cal H}_{\rm el}^0|n>=\epsilon_n |n>$, and the susceptibility
can be computed from
\begin{equation}
\chi_{\ell m}^0 = 2\sum_n^{\rm occ} \sum_{n^{\prime}}^{\rm empty}
\frac{<\ell|n^{\prime}><n^{\prime}|m><m|n><n|\ell>}{\epsilon_{n^{\prime}}
  -\epsilon_n}
\label{eq:chi0}
\end{equation}
However, it is crucial to account for the Coulomb interaction between
electrons, which means solving the electron many-body problem.  This
cannot be done exactly.  The random-phase approximation (RPA, also known
as the time-dependent Hartree approximation) provides good
insight and not a bad numerical answer.  The relevant equation is
\begin{equation}
\chi|\delta V_{\rm ext}> = \chi_0 (|\delta V_{\rm ext}> 
  +|\delta V_{\rm Hartree}>)
\label{eq:rpa}
\end{equation}
Then Eq.(\ref{eq:suscen}) is replaced by
\begin{equation}
{\cal E}_{\rm ch}(\{q_{\ell}\})\approx\frac{1}{2}<q|\chi_0^{-1}+V|q>
\label{eq:rpaen}
\end{equation}
Note that the structure of the classical model Eq.(\ref{eq:elecen})
is closely followed, with the non-interacting susceptibility interpreted as
a non-local capacitance.  The result is an effective dipole-dipole coupling
which replaces the classical Eq.(\ref{eq:gameff}),
\begin{equation}
\Gamma_{\rm eff}= \Gamma + K^{\dagger}[1/\chi_0 +V]^{-1} K
\label{eq:gameffq}
\end{equation}

As an illustration, consider the same
problem of a single polarizable molecule,
with two chargeable sites, one to each side at a distance $r$.
This time the sites have $s$-orbitals coupled by hopping 
matrix element $t$ and Coulomb coupling $q^2/2r$.  It is
also sensible to include an on-site Hubbard repulsion $U$
which discourages deviation from charge neutrality.
This system responds to external fields with the effective
polarizability
\begin{equation}
\frac{1}{\alpha_{\rm eff}}=\frac{1}{\alpha}
-\frac{1/r^4}{t/e^2 +U/2e^2-1/4r}.
\label{eq:alphaeffq}
\end{equation}

Halas {\it et al.} \cite{Halas,Cassagneau} 
have discovered how to coat semiconducting 
nanoparticles with metal shells.  It would be interesting to
attempt this with nanoparticles of large polarizability, in
order to engineer new superparaelectric or nanoferroelectric
systems.

\begin{center}
{\bf Note added in proof, Nov. 15, 2003}
\end{center}

Concerning the possiblity of ferroelectricity
in fullerenes {\it via} the CM polarization
catastrophe, R. L. Whetten had suggested to me endohedral
fullerenes as a route to enhanced polarizability.
It turns out the Clougherty \cite{Clougherty1,Clougherty2}
has speculated about ferroelectricity in such systems,
not from a CM point of view, but more correctly, from
a Jahn-Teller point of view.

N. W. Ashcroft has reminded me that the 
CM polarization catastrophe was formulated by Goldhammer
\cite{Goldhammer} and Herzfeld \cite{Herzfeld}
as a criterion for metallization under
pressure.  As the macroscopic dielectric function
diverges, we now understand that the frequency of some
``ferroelectric soft mode'' goes to zero.  If this
is not a lattice mode, but an electronic mode (plasmon),
then the new phase as understood by Goldhammer and
Herzfeld, should be a metal, not a ferroelectric.  This
criterion proved useful for guiding high pressure
experiments \cite{Ruoff}, particularly for the metallization
of Xe \cite{Goettel,Reichlin}, where there are no lattice modes 
internal to the ``molecule'' of Xe.

N. W. Ashcroft has also reminded me of an interesting extension
of dipole ordering energy theory, namely the
construction of harmonic polarization wave
states, first done by Lundqvist and Sjolander \cite{Lundqvist}.
Atwal and Ashcroft are applying this idea to superconducting
interactions \cite{Atwal}, following earlier work \cite{Ashcroft}.

Finally G. Stell has introduced me to some of the literature on
computer simulations of dipolar systems.  Some
of these papers \cite{Weis1,Weis2,Lavender,Lu1,Lu2,Wolde1,Wolde2,Ayton,Klapp}
are particularly relevant to the discussions in the present paper.

\begin{center}
{\bf Acknowledgments}\\
\end{center}

I thank  S. O'Brien for stimulating this work.  Helpful guidance
came from N. W. Ashcroft, L. E. Brus, J. W. Davenport, R. A. Friesner, 
I. P. Herman, A. J. Millis, and G. Stell.
The work was supported in part by NSF grant no. DMR-0089492, and
in part by a J. S. Guggenheim Foundation fellowship.  Work at
Columbia was supported in part by the MRSEC Program of the NSF
under award no. DMR-0213574.


\begin{thebibliography}{99}

\bibitem{Guyot} M. Shim and P. Guyot-Sionnest, 
  J. Chem. Phys. {\bf 111}, 6955 (1999).

\bibitem{Landauer} R. Landauer, Solid State Commun. {\bf 95}, 7 (1995).

\bibitem{Moro} R. Moro, X. Xu, S. Yin, and W. A. de Heer,
    Science {\bf 300}, 1265 (2003).

\bibitem{Clausius}  R. Clausius, {\it Die Mechanische Warmtheorie} II, 62
Vieweg, Braunschweig (1897).

\bibitem{Mossotti} O.F. Mossotti, Memorie di Mathematica e di Fisica 
della Società Italiana della Scienza Residente in Modena {\bf 24}, 49 (1850).

\bibitem{Aspnes} D. E. Aspnes, Am. J. Physics {\bf 50}, 704 (1982).

\bibitem{Dugourd} Ph. Dugourd, R. Antoine, D. Rayane, I. Compagnon, 
and M. Broyer, J. Chem Phys. {\bf 114}, 1970 (2001).

\bibitem{Rubio} A. Rubio, J. A. Alonso, J. M. Lopez, and M. J. Stott, 
  Phys. Rev. B {\bf 49}, 17397 (1994).

\bibitem{Goddard} A. K. Rappe and W. A. Goddard,
  J. Phys. Chem. {\bf 95}, 3358 (1991).

\bibitem{Friesner} H. A. Stern, G. A. Kaminski, J. L. Banks, R. Zhou,
  B. J. Berne, and R. A. Friesner, J. Phys. Chem. B {\bf 103}, 4730 (1999).

\bibitem{Madden} P. A. Madden and M. Wilson, Chem. Soc. Rev. {\bf 25}, 339
(1996).

\bibitem{Silinsh} E. A. Silinsh and V. Capek, {\it Organic Molecular
   Crystals: Interaction, Localization, and Transport Phenomena} (AIP, 
   New York, 1994).

\bibitem{Rohleder} J. W. Rohleder and R. W. Munn, {\it Magnetism and
   Optics of Molecular Crystals} (Wiley, New York, 1992).

\bibitem{Tsiper} E. V. Tsiper and Z. G. Soos, Phys. Rev. B {\bf 64},
   195124 (2001).

\bibitem{Maksimov} E. G. Maksimov and I. I. Mazin, 
  Solid State Commun. {\bf 27}, 527 (1978).

\bibitem{Ballard}  A. Ballard, K. Bonin, and J. Louderback,
  J. Chem. Phys. {\bf 113}, 5732 (2000).

\bibitem{Pederson} M. R. Pederson and A. A. Quong, 
   Phys. Rev. B {\bf 46}, 13584 (1992).

\bibitem{Eklund}  P. C. Eklund, A. M. Rao, Y. Wang, P. Zhou, K.-A. Wang,
  J. M. Holden, M. S. Dresselhaus, and G. Dresselhaus, Thin Solid Films
  {\bf 257}, 211 (1995).

\bibitem{Jona} F. Jona and G. Shirane, {\it Ferroelectric Crystals},
  Macmillan, New York, 1962.

\bibitem{Lines} M. E. Lines and A. M. Glass, {\it Principles and Applications
  of Ferroelectrics and Related Materials}, Clarendon Press, Oxford, 1977.

\bibitem{Various} This equation appears in the literature in various
forms (see, for example, ref. 8), 
at least as far back as Luttinger and Tisza (next reference).
For a recent version, see L. Jensen, P.-O. Astrand, A. Osted,
J. Kongsted, and K. V. Mikkelsen, J. Chem. Phys. {\bf 116}, 4001 (2002).

\bibitem{Luttinger} J. M. Luttinger and L. Tisza, 
  Phys. Rev. {\bf 70}, 954 (1946).

\bibitem{Fu} H. Fu and L. Bellaiche, in {\it Fundamental Physics
   of Ferroelectrics 2003}, edited by P. K. Davies and D. J. Singh
   (AIP, New York, 2003).

\bibitem{Smith} W. Smith, CCP5 Newsletter No. 46, Oct. 1998, p.18.

\bibitem{Pines} D. Pines, {\it Elementary Excitations in Solids},
W. A. Benjamin, New York, 1964.

\bibitem{White} S. J. White, M. R. Roser, J. Xu, J. T. van der Noordaa,
and L. R. Corruccini, Phys. Rev. Lett. {\bf 71}, 3553 (1993).

\bibitem{Teixeira} P. I. C. Teixeira, J. M. Tevares, and M. M.Telo da Gama,
   J. Phys: Condens. Mat. {\bf 12}, 411 (2000).

\bibitem{Leeuven}  M. E. van Leeuven and B. Smit, Phys. Rev. Lett. 
   {\bf 71}, 3991 (1993).

\bibitem{Gennes} P. G. de Gennes and P. Pincus, Phys. Kondens. Mater.
   {\bf 11}, 189 (1970).

\bibitem{Butter} K. Butter, P. H. H. Bomans, P. M. Frederik,
  G. J. Vroege, and A. P. Philipse, Nature Materials {\bf 2}, 88 (2003).

\bibitem{Liao} J. Liao, Y. Zhang, W. Yu, L. Xu, C. Ge, J. Liu, and N. Gu,
   Colloids and surfaces A {\bf 223}, 177 (2003).

\bibitem{Peebles} L. H. Peebles Jr., {\it Molecular Weight Distribution
in Polymers}, Interscience, New York, 1971.

\bibitem{Madelung} E. Madelung, Phys. Zeits. XIX, 524 (1918).

\bibitem{Watson} R. E. Watson, J. W. Davenport, M. L. Perlman, and T. K. Sham,
   Phys. Rev. B {\bf 24}, 1791 (1981); W. A. Schwalm, Am. J. Phys.
   {\bf 50}, 444 (1982).

\bibitem{Allen} P. B. Allen, Proceedings of the NATO Advanced
Research Workshop ``Molecular Nanowires and Other Quantum Objects,''
to be published by Kluwer, 2004; cond-mat/0309392.

\bibitem{Maschhoff} B. L. Maschhoff and J. P. Cowin, J. Chem. Phys.
   {\bf 101}, 8138 (1994).

\bibitem{Halas} R. D. Averitt, D. Sarkar, and N. J. Halas, 
  Phys. Rev. Lett. {\bf 78}, 4217 (1997); 
  S. J. Oldenberg, R. D. Averitt, S. L. Westcott, and N. J. Halas,  
  Chem. Phys. Lett. {\bf 288}, 243 (1998).

\bibitem{Cassagneau} see also T. Cassagneau and F. Caruso,
   Advanced Mater. (Wenheim) {\bf 14}, 732 (2002) and references therein.

\bibitem{Clougherty1}
D. P. Clougherty, in {\it Fundamental physics of ferroelectrics 2000}, R. E. Cohen,
editor (AIP, Melville, N. Y., 2000) p. 259.

\bibitem{Clougherty2}
D. P. Clougherty, in {\it Vibronic Interactions: Jahn-Teller Effect in
Crystals and Molecules}, M. D. Kaplan and G. O. Zimmerman, Eds. (Kluwer,
Dordrecht, 2001).

\bibitem{Goldhammer} D. A. Goldhammer, {\it Dispersion und Absorption des Lichtes},
B. G. T\"ubner, Leipzig, 1913; p. 27.

\bibitem{Herzfeld} K. F. Herzfeld, Phys. Rev. {\bf 29}, 701 (1929).

\bibitem{Ruoff} A. L. Ruoff, High Pressure Research {\bf 1}, 3 (1988).

\bibitem{Goettel} K. A. Goettel, J. H. Eggert, I. F. Silvera, and
W. C. Moss, Phys. Rev. Lett. {\bf 62}, 665 (1989).

\bibitem{Reichlin} R. Reichlin, Y. K. Vohra, S. Martin, A. K. McMahan, K. E.
Brister, M. Ross, and A. L. Ruoff, Phys. Rev. Lett. {\bf 62}, 669 (1989).

\bibitem{Lundqvist} S. Lundqvist and A. Sjolander, Arkiv for Fysik {\bf 26},
17 (1964).

\bibitem{Atwal} G. S. Atwal and N. W. Ashcroft, manuscript draft (2003).

\bibitem{Ashcroft} N. W. Ashcroft, in {\it Novel Superconductivity}, S. A. Wolf
and V. Z. Kresin, eds.  (Plenum, N. Y., 1987) p. 301; 
and in {\it Recent Progress in Many-Body Theories},
A. J. Kallio, E. Pajanne, and R. F. Bishop, eds. (Plenum, N. Y., 1988) p. 39.

\bibitem{Weis1} J. J. Weis, D. Levesque, and G. J. Zarragoicoechea, Phys. Rev.
Lett. {\bf 69}, 913 (1962). 

\bibitem{Weis2} J. J. Weis and D. Levesque, Phys. Rev. Lett. {\bf 71}, 2729 (1993).

\bibitem{Lavender} H. B. Lavender, K. A. Iyer, and S. J. Singer,
J. Chem. Phys. {\bf 101}, 7856 (1994).

\bibitem{Lu1} D. Lu and S. J. Singer, J. Chem. Phys. {\bf 103}, 1913 (1995).

\bibitem{Lu2} D. Lu and S. J. Singer, J. Chem. Phys. {\bf 105}, 3700 (1996).

\bibitem{Wolde1} P. R. ten Wolde, D. W. Oxtoby, D. Frenkel, Phys. Rev. Lett. {\bf 81}, 3695 (1998).

\bibitem{Wolde2} P. R. ten Wolde, D. W. Oxtoby and D. Frenkel, J. Chem. Phys. {\bf 111}, 4762 (1999).

\bibitem{Ayton} G. Ayton, M. J. P. Gingras, and G. N. Patey , Phys. Rev. Lett.  
{\bf 75}, 2360 (1995).

\bibitem{Klapp} H. L . Klapp and G. N. Patey, J. Chem. Phys. {\bf 115}, 4718 (2001).
     

\end{thebibliography}
\end{document}